# HEAT TRANSFER BY MOBILE LOW-FREQUENCY PHONONS AND "LOCALIZED" MODES IN CRYOCRYSTAL SOLUTIONS.

*V.A. Konstantinov , E.S. Orel, V.P. Reviakin*

*B.I. Verkin Institute for Low Temperature Physics and Engineering, NAS Ukraine, 47, Lenin Ave., Kharkov, 61103, Ukraine*

*e-mail: konstantinov@ilt.kharkov.ua*

## Introduction

The solid inert gases Ar, Kr, and Xe belong to the group of the simplest objects of physics of solids and are therefore used traditionally for comparison of experimental and calculated data [1]. At temperatures close to or above the Debye temperature ($T \geq \Theta_D$) the thermal conductivity of perfect crystals is determined solely by phonon-phonon scattering and it is expected to follow the law $\Lambda \propto 1/T$ [2]. To obey the law, the volume of the crystal should remain invariable, because the modes would otherwise change and so would the temperature dependence of thermal conductivity [2,3].

The isochoric studies of the thermal conductivity of heavy solid inert gases however show a considerable deviation from the above dependence, which is connected with the approach of the thermal conductivity to its lower limit [4,5]. The concept of the lower limit of the thermal conductivity proceeds from the assumption that all the excitations are weakly localized in the regions whose sizes are half the wavelength - $\lambda/2$. As a result, the excitations can hop from site to site through the thermal diffusion [6]. In this case the lower limit of the thermal conductivity $\Lambda_{min}$ of the lattice can be written as:

$$\Lambda_{min} = \frac{1}{2}\left(\frac{\pi}{6}\right)^{1/3} k_B n^{2/3} (v_l + 2v_t), \qquad (1)$$

where $v_l$ and $v_t$ – are the longitudinal and transverse sound velocities, $n = 1/a^3$ - is the number of atoms per unit volume, $k_B$ is the Boltzmann constant.

The further studies of heat transfer in the solid $Kr_{1-\xi}Xe_\xi$ solution (0≤ξ≤0.14) detected a gradual change from the thermal conductivity typical of a perfect crystal to the lower limit of thermal conductivity $\Lambda_{min}$ as the crystal becomes increasingly disordered [7]. In present study the temperature and volume dependences of the thermal conductivity of the solid $Kr_{1-\xi}Xe_\xi$ solution are analyzed in the framework of the model assuming that the phonon mean-free path cannot decrease infinitely.

**Model**

We use Callaway's expression for thermal conductivity

$$\Lambda = \frac{k_B}{2\pi^2 v^2} \int_0^{\omega_D} l(\omega)\omega^2 d\omega, \qquad (2)$$

where v is the sound velocity; $\omega_D$ is the Debye frequency ($\omega_D = (6\pi^2)^{1/3} v/a$); $l(\omega)$ is the phonon mean-free path determined by the U-processes and by point defects scattering:

$$l(\omega) = \left(l_u^{-1}(\omega) + l_i^{-1}(\omega)\right)^{-1}, \qquad (3)$$

The phonon mean-free paths corresponding to each mechanism of scattering are described as [2,9-10]

$$l_u(\omega) = v/A\omega^2 T, \qquad A = \frac{18\pi^3}{\sqrt{2}} \frac{k_B \gamma^2}{\overline{M}\overline{a}^2 \omega_D^3}; \qquad (4)$$

$$l_i(\omega) = v/B\omega^4, \qquad B = \frac{3\pi\Gamma}{2\omega_D^3}; \qquad (5)$$

where the Grüneisen parameter $\gamma = -(\partial \ln\Theta_D/\partial \ln V)_T$, $\overline{M}$ is the average atomic weight of the solution: $\overline{M} = (1-\xi)M_{Kr} + \xi M_{Xe}$; $\overline{a} = (1-\xi)a_{Kr} + \xi a_{Xe}$.

Taking into account the difference ΔM between the atomic (molecular) masses of the impurity and the matrix and the lattice dilatation, the coefficient Γ can be written as

$$\Gamma = \xi(1-\xi)\left(\frac{\Delta M}{\overline{M}} + 6\gamma\frac{\Delta a}{\overline{a}}\right)^2, \qquad (6)$$

where $\Delta M = \overline{M} - M_{Xe}$;  $\Delta a = \overline{a} - a_{Xe}$.

Expression (3) does not apply if $l(\omega)$ becomes of the order of one-half the phonon wavelength $\lambda/2 = \pi v/\omega$ or smaller. A similar situation was considered previously for the case of U-processes alone [10]. Let us assume that in the general case

$$l(\omega) = \begin{cases} v/(A\omega^2 T + B\omega^4) & 0 \leq \omega \leq \omega_0 \\ \alpha\pi v/\omega = \alpha\lambda/2 & \omega_0 < \omega \leq \omega_D \end{cases}, \qquad (7)$$

where $\alpha$ - is a numerical coefficient of the order of unity. There is evidence that the Ioffe-Regel criterion suggesting localization does not hold for phonon gas [11]. Nevertheless, we will refer to the excitations whose frequencies are above the photon mobility edge $\omega_0$ as "localized". Since completely localized states do not contribute to thermal conductivity, we assume that the localization is weak and the excitations can hop from site to site diffusively, as was suggested by Cahill and Pohl [6]. The frequency $\omega_0$ can be found from the condition

$$v/\left(A\omega_0^2 T + B\omega_0^4\right) = \alpha\pi v/\omega_0, \qquad (8)$$

as

$$\omega_0 = \frac{1}{(2\alpha\pi B)^{\frac{1}{3}}}\left[\sqrt[3]{1+\sqrt{1+u}} + \sqrt[3]{1-\sqrt{1+u}}\right], \qquad (9)$$

where the dimensionless parameter u is

$$u = \frac{4\alpha^2 \pi^2 A^3 T^3}{27B}, \qquad (10)$$

If $\omega_0 > \omega_D$, the mean-free path of all the modes exceeds $\lambda/2$, and at $T \geq \Theta_D$ we obtain the well known expression [9]:

$$\Lambda_{ph} = \frac{k_B}{2\pi^2 v}\frac{1}{\sqrt{ATB}}\,\text{arctg}\sqrt{\frac{B}{AT}}\omega_D, \qquad (11)$$

At $\omega_0 \leq \omega_D$ the thermal conductivity integral separates into two parts describing the contributions to heat transfer from the mobile low-frequency phonons and the "localized" high-frequency modes:

$$\Lambda = \Lambda_{ph} + \Lambda_{loc}, \qquad (12)$$

In the high-temperature limit ($T \geq \Theta_D$) these contributions are:

$$\Lambda_{ph} = \frac{k_B}{2\pi^2 v} \frac{1}{\sqrt{ATB}} \text{arctg}\sqrt{\frac{B}{AT}}\omega_0, \qquad (13)$$

$$\Lambda_{loc} = \frac{\alpha k_B}{4\pi v}(\omega_D^2 - \omega_0^2), \qquad (14)$$

The dependence of thermal conductivity on the specific volume is characterized by the Bridgman coefficient [3, 12]:

$$g = -(\partial \ln \Lambda / \partial \ln V)_T, \qquad (15)$$

Taking into account that $(\partial \ln A/\partial \ln V)_T = 3\gamma + 2q - 2/3$, where $q = (\partial \ln \gamma/\partial \ln V)_T$, and $(\partial \ln B/\partial \ln V)_T = 3\gamma$ (as follows from Eqs.(4), (5)), and $(\partial \ln \Gamma/\partial \ln V)_T \approx 0$, we have:

$$g = \frac{\Lambda_{ph}}{\Lambda}g_{ph} + \frac{\Lambda_{loc}}{\Lambda}g_{loc}, \qquad (16)$$

where

$$g_{ph} = -\left(\frac{\partial \ln \Lambda_{ph}}{\partial \ln V}\right)_T = 2\gamma + q + \frac{\sqrt{\frac{B}{AT}}\omega_0}{\left(1 + \frac{B}{AT}\omega_0^2\right)\text{arctg}\sqrt{\frac{B}{AT}}\omega_0}\left(\gamma_0 + q - \frac{1}{3}\right), \qquad (17)$$

$$g_{loc} = -\left(\frac{\partial \ln \Lambda_{loc}}{\partial \ln V}\right)_T = -\gamma + \frac{1}{3} + \frac{2}{\omega_D^2 - \omega_0^2}(\omega_D^2\gamma - \omega_0^2\gamma_0), \qquad (18)$$

$$\gamma_0 = -\left(\frac{\partial \ln \omega_0}{\partial \ln V}\right)_T = \gamma + \frac{u^{\frac{1}{3}}}{6\sqrt{1+u}}\left[\sqrt[3]{1+\sqrt{1+u}} - \sqrt[3]{1-\sqrt{1+u}}\right](6\gamma + 6q - 2); \qquad (19)$$

**Results and discussion**

The isochoric thermal conductivity of the solid $Kr_{1-\xi}Xe_\xi$ ($\xi$=0.034, 0.072, 0.14) solution was studied on samples of different densities in the internal of temperatures from 80K to the onset of melting. The choice of the system, concentrations and temperature interval was dictated by the following.

The phase diagram of the solid $Kr_{1-\xi}Xe_\xi$ solution is well known [13]. The liquid and solid phases have the point of equal concentrations at 114.1K and $\xi$=0.15. Between 75 and 114K the component form a solid fcc solution for all $1 \geq \xi \geq 0$. When samples with the temperature gradient along the measuring cell are grown, the solid solution can become layered at $\xi > 0.15$. The highest Xe concentration was therefore limited to 14%.

The isochoric thermal conductivities of pure Kr and the $Kr_{1-\xi}Xe_\xi$ solution for which the isochoric condition comes into play at 80K are shown in Figs. 1-4 (black squares). The computer fitting of thermal conductivity using Eqs. (12) – (14) was performed by the least square method varying the coefficients A and $\alpha$. The parameters of the Debye model for thermal conductivity used in the fitting (a, v [1, 13]; $\Gamma$ - coefficients calculated by Eq. (6)) and the fitted values A and $\alpha$ are listed in Table 1 along with the Bridgman coefficients obtained in the experiment and calculated by Eq. (16)-(19). The calculation was made using values $\gamma = 2.5$ and $q = 1$ [1,13].

The fitting results are shown in Figs. 1-4 (solid lines). The same figures show the contributions (dash-dot lines) to the thermal conductivity from the mobile low-frequency phonons - $\Lambda_{ph}$ and from the "localized" high-frequency modes - $\Lambda_{loc}$. The dashed line in the figures indicate the lower limits of thermal conductivity $\Lambda_{min}$, which were obtained as asymptotes of the $\Lambda(T)$ dependence at $T \to \infty$.

It is seen in Fig. 1 that in pure Kr the "localization" of the high-frequency modes starts above 90K. As the temperature rises, the amount of heat transferred by the "localized" modes increases and at 160 K it becomes equal to the heat transferred by the mobile low-frequency phonons. In the solution with $\xi = 0.034$ (see Fig. 2) the "localization" of the high-frequency modes starts at 30 K and above 100 K most of the heat is transferred by the "localized" modes. As the temperature and the impurity concentration increase, so does the amount of heat transferred by the "localized" modes. With $\xi = 0.14$ (see Fig. 4) practically all the heat at T≥$\Theta_D$ is transferred by the "localized" modes. The lower limit of thermal conductivity found by fitting is 1.1-1.2 times higher than $\Lambda_{min}$ calculated by Eq. (1).

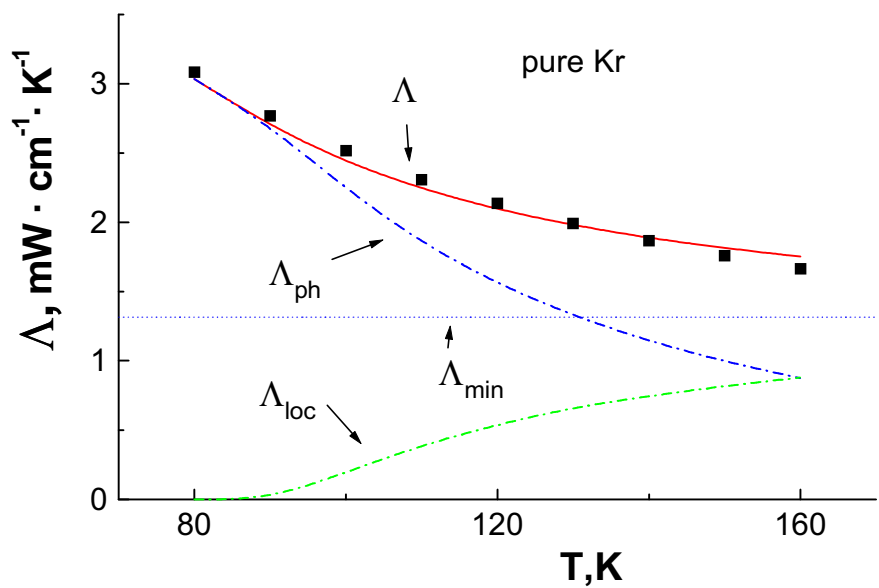

Fig. 1. Fitting results for isochoric thermal conductivity and calculated relative contributions of mobile low-frequency phonons and "localized" modes to the thermal conductivity of pure Kr (molar volume is 28,5 см³/mol).

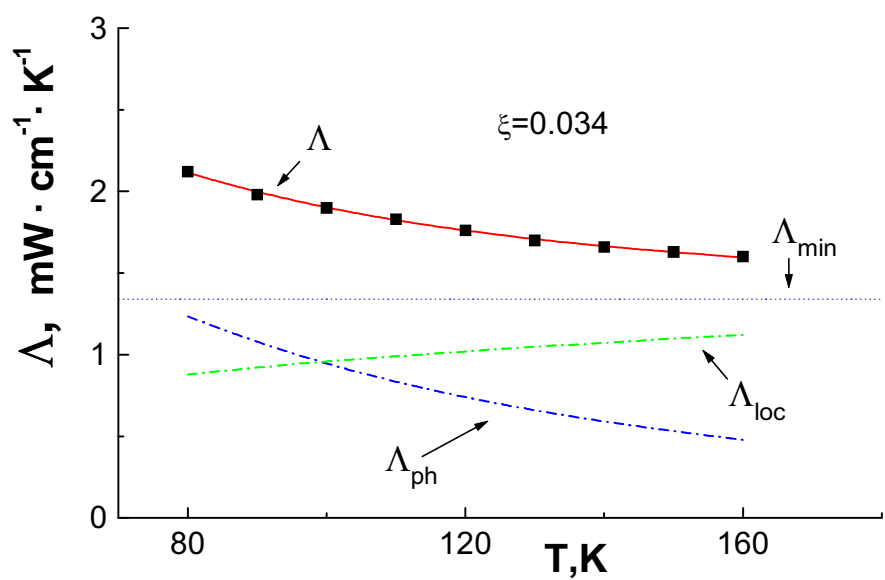

Fig. 2. Fitting results for isochoric thermal conductivity and calculated relative contributions of mobile low-frequency phonons and "localized" modes to the thermal conductivity of solid $Kr_{1-\xi}Xe_\xi$ ($\xi=0.034$) solution (molar volume is 29.1 см³/mol).

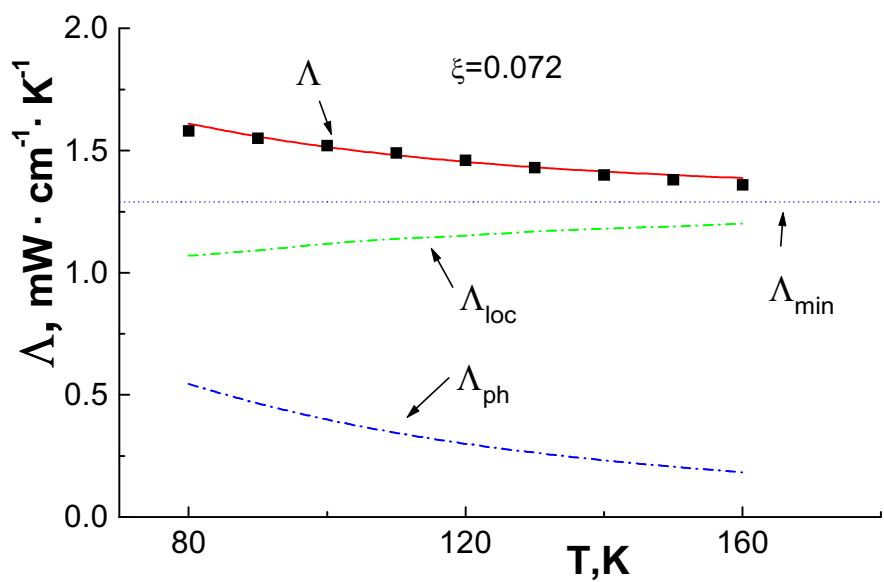

Fig. 3. Fitting results for isochoric thermal conductivity and calculated relative contributions of mobile low-frequency phonons and "localized" modes to the thermal conductivity of solid $Kr_{1-\xi}Xe_\xi$ ($\xi=0.072$) solution (molar volume is 29.4 см$^3$/mol).

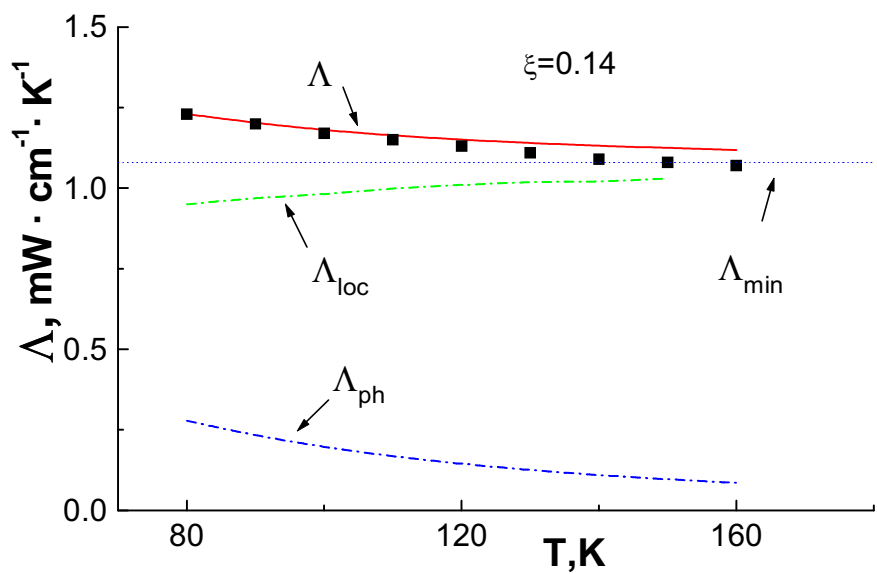

Fig. 4. Fitting results for isochoric thermal conductivity and calculated relative contributions of mobile low-frequency phonons and "localized" modes to the thermal conductivity of solid $Kr_{1-\xi}Xe_\xi$ ($\xi=0.14$) solution (molar volume is 29.8 см$^3$/mol).

As seen in Table 1, the experimental and calculated Bridgman coefficients are in fairly good agreement if it is remembered that the g-value is estimated with large uncertainty and the model disregards phonon dispersion and the real density of states. The temperature dependence of the Bridgman coefficients $g = -(\partial \ln \Lambda / \partial \ln V)_T$ of solid $Kr_{1-\xi}Xe_\xi$ solution calculated by Eqs. (16)-(19) is shown in Fig. 5. Eqs. (16) - (19) describe the general tendency of the Bridgman coefficient g to decrease as the crystal becomes increasingly disordered and most of the heat transferred by the "localized" modes.

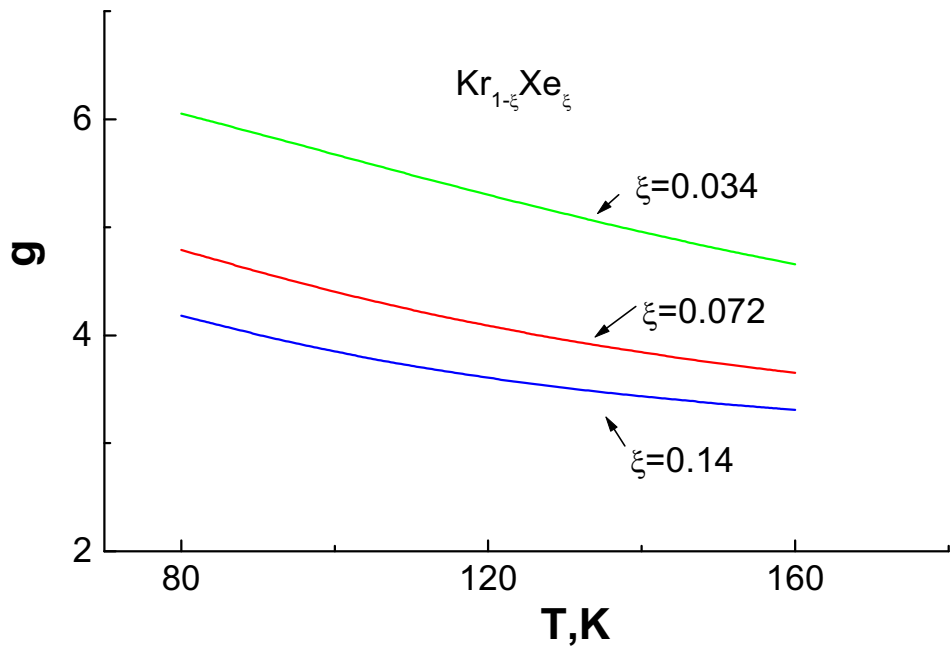

Fig. 5. Temperature dependence of the Bridgman coefficient $g = -(\partial \ln \Lambda / \partial \ln V)_T$ for solid $Kr_{1-\xi}Xe_\xi$ solution.

Table 1. Parameters of the Debye model for thermal conductivity used in fitting: a, v and $\Gamma$; obtained through fitting A and $\alpha$; calculated $g_{th}$ and experimental $g_{exp}$ the Bridgman coefficient.

| $\xi$ | $a \times 10^{-8}$ cm | v km/sec | $\Gamma$ | $A \times 10^{-16}$ sec/ K | $\alpha$ | $g_{exp}$ | $G_{th}$ |
|---|---|---|---|---|---|---|---|
| 0 | 3.62 | 0.86 | 0 | 3.1 | 1.2 | 9.4 | 9.2 |
| 0.034 | 3.64 | 0.86 | 0.1 | 3.8 | 1.2 | 8.0 | 5.7 |
| 0.072 | 3.65 | 0.87 | 0.19 | 6.4 | 1.2 | 5.5 | 4.6 |
| 0.14 | 3.67 | 0.87 | 0.29 | 9.8 | 1.1 | 4.0 | 3.8 |

**Conclusions**

It is shown that the temperature and volume dependences of the thermal conductivity of the solid $Kr_{1-\xi} Xe_{\xi}$ ($\xi < 0.14$) solution can be described in the framework of the model in which heat is transferred by mobile low-frequency phonons; above the phonon mobility edge, heat is transferred by the "localized" modes migrating randomly from site to site. The phonon mobility edge $\omega_0$ is found from the condition that the phonon mean-free path determined by the Umklapp processes and scattering on point defects cannot become smaller than one-half the phonon wavelength. The Bridgman coefficients $g = -(\partial \ln \Lambda / \partial \ln V)_T$ is the weighted-mean over these modes differing widely in their volume dependence. It is shown that the amount of heat transferred by the "localized" modes is quite large above 100 K even in pure Kr and it increases with rising temperature and impurity concentration.

The authors are indebted to V.G. Manzhelii, Full Member of NAS of Ukraine, and Prof. R.O. Pohl for fruitful discussions.

**References**

1. Rare gas solids, v. I-II, M.L. Klein & J.A. Venables (eds.), Academic Press, London, New York, (1977).
2. R. Berman, Thermal Conduction in Solids, Oxford, Clarendon Press (1976).
3. G.A. Slack, in: Solid State Physics, H. Ehrenreich, F. Seitz and D. Turnbull (eds.) Academic Press, New York, London, **34**, 1 (1979).
4. V.A. Konstantinov, V.G. Manzhelii, M.A. Strzhemechnyi, S.A. Smirnov, LTP, **14**, 48 (1998).
5. V.A. Konstantinov, JLTP, **122**, 459 (2001).
6. D.G. Cahill, S.K. Watson and R.O. Pohl, Phys. Rev. B, **46**, 6131 (1992).
7. V.A. Konstantinov, R.O. Pohl, V.P. Revyakin, LTP, **44**, 857 (2002).
8. J. Callaway, Phys. Rev. **113**, 1046 (1959).
9. P.G. Klemens, High. Temp. High Pressure, **5**, 249 (1983).
10. M.C. Roufosse, and P.G. Klemens, J.Geophys. Res. **79**, 703 (1974).
11. J.L. Feldman, M.D. Kluge, P.B. Allen, F. Wooten, Phys. Rev. B., **48**, 12589 (1993).
12. R.G. Ross, P.A. Andersson, B. Sundqvist, and G. Backstrom, Rep. Prog. Phys. **47**, 1347 (1984).
13. V.G. Manzhelii, A.I. Prokhvatilov, I.Ya. Minchina, and L.D. Yantsevich, Handbook of Binary Solutions of Cryocrystals, Begell House Inc., NY (1996).